\def\year{2023}\relax
\documentclass[letterpaper]{article} 
\usepackage{aaai22}  
\usepackage{times}  
\usepackage{helvet}  
\usepackage{courier}  
\usepackage[hyphens]{url}  
\usepackage{graphicx} 
\usepackage{natbib}  
\usepackage{caption} 
\DeclareCaptionStyle{ruled}{labelfont=normalfont,labelsep=colon,strut=off} 
\usepackage{subcaption}
\usepackage{tabularx}
\usepackage{booktabs}
\usepackage{comment}
\usepackage{multirow}
\usepackage{amsmath}
\usepackage{soul}
\usepackage{xspace}
\usepackage{xcolor,colortbl}

\usepackage{xcolor}

\title{Watch Your Language: Investigating Content Moderation with Large Language Models}
\author {
    Deepak Kumar,\textsuperscript{\rm 1, 2} 
    Yousef AbuHashem,\textsuperscript{\rm 1}
    Zakir Durumeric\textsuperscript{\rm 1}\\
}
\affiliations{

    \textsuperscript{\rm 1}Stanford University,
    \textsuperscript{\rm 2}UC San Diego
  }

\begin{document}
\nocopyright
\maketitle

\begin{abstract}
Large language models (LLMs) have exploded in popularity due to their ability to perform a wide array of natural language tasks. Text-based content moderation is one LLM use case that has received recent enthusiasm, however, there is little research investigating how LLMs perform in content moderation settings. In this work, we evaluate a suite of commodity LLMs on two common content moderation tasks: rule-based community moderation and toxic content detection. For rule-based community moderation, we instantiate 95~subcommunity specific LLMs by prompting GPT-3.5 with rules from 95~Reddit subcommunities. We find that GPT-3.5 is effective at rule-based moderation for many communities, achieving a median accuracy of 64\% and a median precision of 83\%. For toxicity detection, we evaluate a suite of commodity LLMs (GPT-3, GPT-3.5, GPT-4, Gemini Pro, LLAMA 2) and show that LLMs significantly outperform currently widespread toxicity classifiers. However, recent increases in model size add only marginal benefit to toxicity detection, suggesting a potential performance plateau for LLMs on toxicity detection tasks. We conclude by outlining avenues for future work in studying LLMs and content moderation.
\end{abstract}

\section{Introduction}
\noindent\fbox{%
    \parbox{0.96\columnwidth}{%
        \textbf{Content Warning}: When
        necessary for clarity, this paper directly quotes user content
        that contains offensive/hateful speech, profanity, and other potentially
        triggering content.
    }%
}
\\\\
Since the release of OpenAI's ChatGPT in November 2022, publicly available large language models (LLMs) have exploded in popularity and achieved widespread recognition. Such systems, like OpenAI's GPT-suite and Google's Gemini, offer new and accessible ways for end-users to interface with AI as well as provide developers with a new toolkit with which to build AI-powered technologies. Indeed, developers are already using LLMs for a variety of text-based tasks: Microsoft incorporated GPT-4 into Bing to enhance web search functionality,\footnote{\url{https://www.theverge.com/2023/3/15/23641683/microsoft-bing-ai-gpt-4-chatbot-available-no-waitlist}} Expedia released a GPT-3.5 powered chatbot to help end-users create travel plans,\footnote{\url{https://mashable.com/article/expedia-chatgpt-in-app-travel-booking}} and Slack incorporated a variety of LLMs to provide summaries of meetings and increase end-user productivity.\footnote{\url{https://slack.com/blog/news/introducing-slack-gpt}}

One use case for LLMs that has achieved significant attention is automated content moderation. Although using AI for content moderation is not a new idea~\cite{gillespie2020content}, recent research has begun investigating how best to incorporate LLMs into moderation pipelines~\cite{franco2023analyzing} and industry players alike are expressing enthusiasm for LLM-powered moderation~\cite{gilardi2023chatgpt}. For example, OpenAI recently published an article outlining how GPT-4 could be used to build and evaluate content policies with test examples\footnote{\url{https://openai.com/blog/using-gpt-4-for-content-moderation}} to minimize time required to develop platform policy and reduce emotional burden for human reviewers. Given this recent attention, it behooves the research community to study the role that LLMs might play in automated content moderation, both in where LLMs excel and where there are opportunities for improvement.

In this paper, we evaluate the effectiveness of several publicly available LLMs on two common content moderation tasks: rule-based community moderation and toxic content detection. For each task, we develop a test harness that combines real-world datasets with careful prompt design and evaluate each model against available ground truth. We focus on OpenAI's GPT models (3, 3.5, 4), Google's Gemini Pro, and Meta's LLAMA as recent exemplars of LLMs deployed in practice for our evaluation. For the rule-based community moderation task, we instantiate 95~subcommunity specific LLMs by prompting GPT-3.5 with community guidelines sourced form 95~subcommunities on Reddit. We then prompt each subreddit-specific model to simulate moderation decisions on real comments sourced from each community and compare how well LLM decisions align with human-moderator decisions. For toxic content detection, we construct a balanced dataset of 10K~comments sourced by Kumar et~al.~\shortcite{kumar2021designing} and compare GPT-3, GPT-3.5, GPT-4, Gemini Pro, and LLAMA 2 performance with ground truth and other popular state-of-the-art commercially available baselines.

We observe that GPT-3.5 is effective at conducting rule-based moderation, achieving an accuracy of 64\% and precision of 83\% for the median subreddit. For some communities, GPT-3.5 achieves near human-moderator levels of performance: for example, the model achieves an accuracy of 82\% and a precision of 95\% on comments from \texttt{r/movies}. We investigate what rules GPT-3.5 evokes when simulating moderation decisions, and find that the model is more likely to moderate based on restrictive rules (e.g., those that restrict behavior) compared to prescriptive rules or formatting based rules, suggesting that framing policies as restrictions may be an effective strategy for moderators and policymakers seeking to use LLMs for moderation tasks. We also highlight potential pitfalls with using LLMs for moderation: we observe stark differences in performance when underlying models change over time, which can lead to unpredictable results.

With toxicity detection, we observe that modern LLMs outperform existing commercial toxicity detection classifiers, with GPT-3.5 achieving the most balanced performance across all tested models (Accuracy of 0.73, F1 of 0.75). Interestingly, we find only marginal improvement in toxicity detection performance when comparing GPT-3, GPT-3.5, GPT-4, and Gemini Pro, suggesting that increased LLM-size may not always lead to increased benefits in a content moderation context. Still, LLMs are not perfect, and exhibit some significant issues: models often miss implicit toxicity, which has been a topic of recent study in toxicity detection~\cite{elsherief2021latent}, and, like existing classification-based approaches, at times cannot make accurate decisions without appropriate world context.

We view our results as a tempered by optimistic first step in introducing LLMs into content moderation contexts. LLMs provide some improvements over existing baselines but will still require human review and careful guardrails before widespread adoption. We conclude with a discussion of our results and chart a pathway for future research in this space. We will make our code and data available at publication time; data is available anonymously with this submission.\footnote{\url{https://anonymous.4open.science/r/llm-content-mod-1F4B}} We hope our results will serve useful to researchers and practitioners investigating how best to incorporate LLMs into content moderation pipelines.
\section{Background and Related Work}
Recent research has investigated the role that machine learning and AI-assisted tools can play in tackling content moderation problems. In this section, we provide the necessary background and detail the relevant research we build on to conduct our analyses.

\subsection{AI-Assisted Content Moderation}
As machine learning and AI tools have become ubiquitous, recent research has investigated the role that such tools might play in content moderation. One class of work has examined the tensions and affordances of adding machine learning to content moderation workflows, demonstrating that while automation can enable new ways of tackling content moderation (e.g., through delegation, scale, third-party tooling), in many cases, human moderators still want to make decisions that fall into gray areas~\cite{kuo2023unsung,ma2023defaulting,atreja2023appealmod,lai2022human,jiang2023trade,han2023hate}. Other work has focused on the design space of adding AI systems to content moderation tasks, particularly through the lens of transparency in moderation decisions~\cite{ma2023defaulting,choi2023creator,jhaver2019does}. 

These studies demonstrate how transparency through explanation often leads to prosocial outcomes in online communities, but challenges with scale often preclude such explanations from widespread adoption. Finally, recent systems-focused work examined how to build fully-fledged AI systems that support existing content moderation needs. Chandrasekharan et~al.\ proposed Crossmod~\shortcite{chandrasekharan2019crossmod}, a system that surfaced potential candidates for moderation to community moderators on Reddit based on outcomes in other subcommunities, and showed significant alignment between the AI system and human-moderator decisions. Our work builds on this prior research by examining how well LLMs fare at both enforcing written community guidelines as compared to human moderators, and examining if LLMs can provide useful explanations of its moderation outcomes.

\subsection{Detecting Toxic Content}
Google Jigsaw's Perspective API~\cite{perspectiveapi} has emerged as the de facto classifer for detecting toxic content online, and it has been used in dozens of research papers to identify toxic content in isolation~\cite{xia2020exploring, kumar2021designing,lambert2022conversational,kumar2023understanding}. There has been significant additional research on improving detection beyond the Perspective API\@. Hanley et~al., for example, demonstrate how a DeBERTa classifier trained with contrastive learning marginally outperforms the Perspective API~\shortcite{hanley2023twits}, and, in a forthcoming paper, He et~al.\ show how prompting local LLMs (e.g., T5, GPT2-M and GPT2-L) can improve performance for a variety of toxicity related tasks~\shortcite{he2024prompt}. Multiple studies have considered how older LLMs (i.e., GPT-3) can be used to identify hate speech~\cite{chiu2021detecting, schick2021self}, and recent research has studied how effective GPT-3.5 is at detecting hate speech in English~\cite{li2023hot} and Portugese~\cite{oliveira2023good}.
\section{Rule-Based Content Moderation}
In this section, we study how LLMs can aid in rule-based moderation settings by instantiating 95~subreddit-specific LLMs with lists of community guidelines sourced from 95~subreddits. We ask each subreddit-specific model to simulate moderation decisions for comments sourced from each subreddit and evaluate their performance.

\subsection{Methods}
\subsubsection{Collecting ground-truth moderation decisions}
To collect ground-truth moderation decisions, we leverage a dataset created by Chandrasekharan et~al.~\shortcite{chandrasekharan2018internet} that contains comments that were moderated from 100~popular subreddits between 2016--2017. Since their study was conducted, 5~subreddits have gone offline or been removed from the platform, so our study focuses on the remaining 95~subreddits. For each subreddit, we sample 500~comments that were removed by moderators as well as 500~comments that were \textit{not} removed by moderators during the same time period, thus creating a \textit{balanced dataset} of moderation decisions for each subreddit. These datasets serve as our baseline for rule-based evaluation.

\subsubsection{Extracting Subreddit metadata}
We first extract the description of each subreddit using the Reddit API. We then extract the moderation rules of that community by using the Reddit rules API,\footnote{https://www.reddit.com/dev/api/\#GET\_r\_subreddit\_about\_rules} which provides a list of rule texts and whether the rule applies to a post, comment, or all content posted to the subreddit. Given we are only investigating comment moderation in this study, we filter any rules that do not apply to comments. We extract the full rule text for each rule and build a numbered list of rules that we use to prompt each model. 

\subsubsection{Prompt design}
We instantiate a new zero-shot prompted LLM for each subreddit we investigate. We begin by prompting the LLM with the description of the subreddit as well as the constructed rules string for each subcommunity. The full initial prompt text is:

\begin{small}
\begin{verbatim}
You are a helpful content moderation 
assistant for the online subreddit 
<subreddit>. The subreddit has the 
following description: <description>.

Here are the rules for the subreddit:
<rule_texts>
\end{verbatim}
\end{small}

We then ask the model to evaluate whether it would moderate new unseen comments (yes/no), the rule text and rule numbers that the comment violates, and an explanation for the decision. We ask the model to return a JSON object with consistent indexing for ease of downstream analysis. The full prompt text is:

\begin{small}
\begin{verbatim}
Consider the following comment:
<example_comment>

Return a JSON object with five fields, 
"would_moderate," that is 
either "yes" or "no" depending 
on if you would remove this comment 
from the subreddit, "rules" which are 
the text of the rules being violated, 
"rule_nums" which are a comma-separated
list of rules being violated, "rating"
which is a score from 1-5 on how 
violative the comment is, and "explanation"
which provides a reason for your decision.
\end{verbatim}
\end{small}

\subsubsection{Evaluation strategy}
We compare ground-truth moderation decisions with decisions made by each subreddit-specific LLM on a number of binary performance metrics. For this experiment, we solely evaluate OpenAI's GPT-3.5 as an exemplar LLM as it balances at-scale measurements (i.e.,~100,000 queries in total) with cost. We set the ``temperature'' to 0 to ensure our results are deterministic across repeated queries. We also conduct a case study comparison between with Google's Gemini Pro for a small number of subreddits to examine if our results generalize to other LLMs. To understand \textit{why} each subreddit-specific model makes mistakes, we also perform a case study on mistakes created in the \texttt{r/worldnews} subreddit and provide a qualitative analysis of existing gaps in rule-based moderation.

\begin{figure}
    \includegraphics[width=\columnwidth]{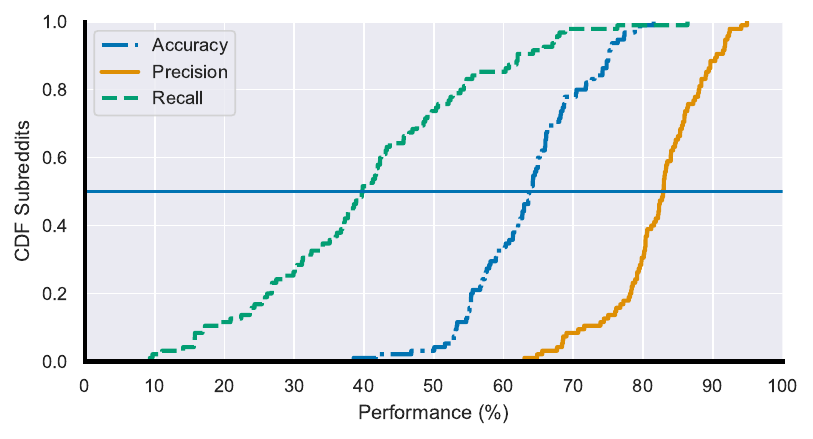}
    \caption{\textbf{Rule-based moderation aggregate performance by Subreddit}---%
        We show a CDF of accuracy, precision, and recall across the 95~subreddits we study. There is considerable spread in performance across subreddits, with GPT-3.5 performing worse than random chance on some and performing near identically to human moderators for others. 
    }
    \label{fig:rule_accuracy}
\end{figure}

\subsection{Aggregate Performance}
To begin, we measure the binary performance of each subreddit-specific LLM when compared to human moderators. GPT-3.5 achieves a median accuracy of 63.7\%, a median precision of 83\%, and a median recall of 39.8\% across all subreddits in our dataset (Figure~\ref{fig:rule_accuracy}). For each metric, there is considerable spread: GPT-3.5 achieves high accuracy for subreddits like \texttt{r/movies} (81.5\%) and \texttt{r/OldSchoolCool} (76.3\%), with each of these also exhibiting high precision (0.95 and 0.82, respectively). Such performance is not uniform across all subreddits. GPT-3.5 performs \textit{worse than a coin toss} when simulating decisions from subreddits like \texttt{r/askscience} (Accuracy 38.5\%) and \texttt{r/AskHistorians} (Accuracy 41.7\%)---suggesting that LLMs are not always a useful community moderation-aid. Notably, both of these subreddits are centered around discussion with experts, and moderation decisions in these subcommunities often require significant world context, which is a limitation of our evaluation setup~\cite{gilbert2020run}.



\subsection{Prompt Sensitivity}
\begin{table}[ht]
    \centering
    \small
    \begin{tabularx}{\columnwidth}{Xrrrr}
    \toprule
    Subreddit   &   Metrics &   Base    &   Platform Rules     &   CoT \\
    \midrule
    \multirow{3}{*}{r/movies}   &   Accuracy & 0.81 & \textbf{0.82} & 0.81 \\
                                &   Precision & \textbf{0.95} & 0.92 & 0.92 \\
                                &   Recall & 0.68 & \textbf{0.72} & 0.7 \\
                                &   F1 & 0.79 & \textbf{0.81} & 0.8 \\
    \midrule
    \multirow{3}{*}{r/IAmA}     & Accuracy & \textbf{0.64} & 0.64 & 0.62 \\
                                & Precision & \textbf{0.84} & 0.82 & 0.78 \\
                                & Recall & 0.37 & \textbf{0.4} & 0.38 \\
                                & F1 & 0.52 & \textbf{0.53} & 0.51 \\
    \midrule
    \multirow{3}{*}{r/askscience}   &   Accuracy & 0.38 & 0.39 & \textbf{0.4} \\
                                    &   Precision & 0.87 & \textbf{0.95} & \textbf{0.95} \\
                                    &   Recall & \textbf{0.11} & 0.11 & 0.11 \\
                                    &   F1 & \textbf{0.19} & 0.19 & 0.19 \\

    \bottomrule    
    \end{tabularx}
    \caption{\textbf{Rule-based moderation prompt sensitivity}---%
        Rule-based content moderation is robust to differing prompt strategies; overall performance is 
        comparable across the three strategies we tested for each subreddit. 
    }
    \label{table:prompt_strategies_reddit}
\end{table}

LLM output is often highly sensitive to input prompts~\cite{white2023prompt,openai-prompt-guides}. As such, we also examine how our results vary with different prompting strategies. We evaluate two prompt strategies in addition to our base prompt: providing more world-context by embedding platform rules, and chain-of-thought prompting, which is known to elicit reasoning and often better performance from commodity LLMs~\cite{wei2022chain}. For the embedded world-context prompt, we prepend to the community rules a list of comment-related rules from the Reddiquette\footnote{https://support.reddithelp.com/hc/en-us/articles/205926439-Reddiquette}, which is a list of platform-wide rules that most subreddits adhere to. For chain-of-thought prompting, we include the phrase ``Let's think step by step.'' at the end of our prompt. For this experiment, we evaluate comments from three subreddits: \texttt{r/movies}, \texttt{r/IAmA}, and \texttt{r/askscience}; these represent communities where GPT-3.5 performed the best, median, and worst in our dataset. We select a subsample of subreddits as the cost of running every subreddit query with every prompt variant was prohibitively expensive (a single run costs upwards of \$200 US dollars).

In general, our performance results are robust to different prompt strategies (Table~\ref{table:prompt_strategies_reddit}). The inclusion of the Reddiquette and CoT-prompting does marginally shift the tradeoff made between precision and recall: in general, embedding the Reddiquette or using CoT prompting resulted in slightly lower precision and slightly higher recall when compared to the base prompt, however, aggregate performance is near identical between the three. These observations are flipped when moderating \texttt{r/askscience}---including platform rules and CoT prompting increased precision meaningfully (from 0.87 to 0.95) without any change to recall, suggesting that for some communities, adding context to prompts can provide significant utility. 



\subsection{Errors in Rule-Based Moderation}
When GPT-3.5 misaligns with human moderators, it is much more likely to create a false negative (86.9\% of all errors) than a false positive (13.1\% of all errors). The error distribution for the median subreddit is similar, with 12.1\% of errors being false positives and 87.8\% of errors being false negatives. Similarly to our prior results, there is significant distributional spread and several subcommunities buck this trend: for example, when moderating the subreddit \texttt{r/NSFW\_GIF}, an adult community for sharing not-safe-for-work imagery, GPT-3.5 creates 68\% false positives and 31\% false negatives, often due to missing context or the model erroneously flagging sexual content, which is desired in this community. These results suggest that while GPT-3.5 favors precision over recall in aggregate, these results do not uniformly impact all subcommunities and depend on content, context, and community norms.

\begin{table}
    \centering
    \begin{tabularx}{\columnwidth}{Xrr}
    \toprule
                &   GPT-3.5/030123  &   GPT-3.5/061323 \\
    \midrule
    Accuracy    &   68\%  & 63\% \\
    Precision   &   75\%    & 83\% \\
    Recall      &   63\%    & 40\% \\
    F1          &   68\%    & 53\%  \\
    \bottomrule        
    \end{tabularx}
    \caption{\textbf{Longitudinal LLM performance}---%
        We show the median performance of GPT-3.5 across the first and second released models (from March 1, 2023 to June 13, 2023).  We find that GPT-3.5 produces drastically different results, with new models significantly preferring precision over recall. Our results suggest that human moderators need to stay significantly in the loop as models grow and evolve.
    }
    \label{table:longitudinal_performance}
\end{table}

\subsection{Stability of LLM Performance Over Time}
Given systems like GPT-3.5 are constantly undergoing changes due to model updates and deployment changes, we also sought to understand how stable model performance is at rule-based moderation tasks over time. Table~\ref{table:longitudinal_performance} shows the median performance across all subreddits in our dataset from the first released GPT-3.5 model (released on March 1, 2023) and the second released model, which is the one we report on in this paper (released on June 13, 2023). Performance significantly changes between the two models, despite there being no other differences in prompts or analyzed comments. F1 performance decreased from 0.68 to 0.53, primarily due to a decrease in median recall: which similarly decreased from 0.63 to 0.40. Median precision, however, increased from 0.75 to 0.83---suggesting that the newer model favored reducing false positives over reducing false negatives. Although it is unclear \textit{why} the underlying GPT-3.5 model shifted in this direction, our results highlight that content moderators wishing to incorporate LLMs into their moderation workflows will need to consider model stability as an additional factor when deploying these tools.


\subsection{LLMs and Rule Enforcement}
LLM decisions are not only governed by the content, but also by the \textit{types of rules} that are available when making moderation decisions. As part of our prompt design, we ask the LLM to express which rules are violated when it flags a comment for moderation. We use these rules to study what kinds of rules LLMs may be better at expressing over other rules, which can inform how LLMs can be better used in practice by community moderators. To increase our descriptive power for each rule, we replicate the work of Fiesler et~al.~\shortcite{fiesler2018reddit}, who created a model and taxonomy for classifying Reddit rules into broad categories. We classify each subreddit rule into three categories: restrictive rules (i.e., those that restrict types of behavior), prescriptive rules (i.e., those that tell someone what to do), and format related rules. We detail our model replication process in the Appendix. 


GPT-3.5 expresses a significant fraction of rules: of the 562~rules in our dataset, GPT-3.5 expressed 467 (83.1\%) when making moderation decisions. The median subreddit has 88\% of its rules expressed by GPT-3.5, with 49\% of subreddits having 100\% of their rules expressed by the LLM. Still, there is significant distributional spread, with some subreddits having only a small fraction of their rules expressed. For example, \texttt{r/AskTrumpSupporters}, a subreddit for asking questions to those that supported Donald Trump's presidency, only has 3~of its 7~rules expressed.

\begin{table}
    \centering
    \small
    \begin{tabularx}{\columnwidth}{Xrrr}
    \toprule
    Rule Type   &   \% Overall Rules    &   \% Subreddits   &   \% Expressed \\
    \midrule
    Restrictive &  460 (82.4\%) & 90 (95\%)  & 398 (87\%)  \\
    Prescriptive    &   319 (57.1\%) & 86 (90.1\%)  & 246 (77\%)  \\
    Format  &  46 (8.2\%) &  28 (29\%) &  33 (71\%) \\
    \bottomrule
    \end{tabularx}
    \caption{\textbf{Rule type expression}---%
        We show the distribution of rule types across rules and subreddits, as well as how frequently each type was expressed by the LLM when making moderation decisions. LLMs are most likely to moderate via restrictive rules (87\%) compared to prescriptive rules (77\%) and format-driven rules (71\%).
    }
    \label{table:rule_types}
\end{table}

Despite high rule coverage overall, GPT-3.5 does not express all rule types uniformly. Table~\ref{table:rule_types} shows the distribution of rule-types in our dataset, across all rules, the fraction of subreddits that contain at least one rule of each type, and the fraction of rules expressed for each rule type. GPT-3.5 was much more likely to enforce restrictive rules (87\%) than prescriptive rules (77\%) and format-related rules (71\%); in both cases, the LLM is statistically significantly less likely to express either rule type when compared to a restrictive rule per a 2-tailed proportions test ($p<0.01$). Our results suggest that GPT-3.5 are more likely to take action for restrictive rules, suggesting that policymakers wishing to use LLMs for rule-enforcement might need to consider more restrictive framings of rules versus prescriptive framings.

Finally, we investigate if errors made by the LLM in moderation decisions are imbalanced across rule types. Understanding this can illuminate whether LLMs might be better suited to moderate certain types of rules versus other rules. Given that the LLM only expresses a rule if it chooses to moderate the comment, we measure the error rate for each rule type by the fraction of comments for each rule type that the LLM made a false positive. Surprisingly, we find little evidence that rule types impact performance: we find almost no difference in error rates for restrictive rules (18.5\% errors), prescriptive rules (15.9\% errors) or formatting rules (16.9\% errors), highlighting that GPT-3.5 erroneously flags content for moderation across all rule types evenly.

\subsection{Moderating Subreddits in Realtime}
Our analysis thus far has focused on evaluating LLMs in a \textit{balanced class} scenario to provide a benchmark for performance. To understand how LLMs perform in a more \textit{realistic} scenario, we applied our subreddit-specific LLMs to realtime decisions made by community moderators. To capture this, we collected our own realtime dataset by repeating the procedure in Chandrasekharan et.~al~\cite{chandrasekharan2018internet}. We first stream comments from the 95~subreddits of interest, and second, query which of those comments were eventually removed (denoted with the text of the comment changing to ``[removed]'') by moderators after 24-hours. We collected realtime moderation decisions for a 3~week period between November~11 2023 and and December~18 2023.

Given the cost constraints with querying GPT-3.5, we did not run every comment through each subreddit-specific LLM. Rather, we sampled moderated and unmoderated comments from each subreddit at each subreddit's moderation rate. In general, moderation rates were low, with an average of 2\% of comments moderated and rates ranging from as small as 0.2\% (\texttt{r/CFB}) to at most 15.7\% (\texttt{r/legaladvice}). We selected a sample size for each subreddit large enough to ensure our results are robust at an error rate of 5\% and a CI of 95\%, and did not analyze any subreddits for which we were not able to achieve sufficient comment volume. Ultimately, we curated and analyzed realtime moderation decisions from 62~of our 95~subreddits.

The realtime moderation task is more challenging for each subreddit-specific LLM than a balanced task, with the average ROC\_AUC decreasing from 0.67 to 0.6 across all comments. Some subreddits, however, saw \textit{greater} performance in a realtime setting: for example, the ROC\_AUC for \texttt{r/gonewild} increased from 0.56 to 0.61. Accuracy for the realtime task increases from 70\% to 90\%, however, this is largely due to a significant increase in unmoderated comments with significant class imbalance. Average recall drops from 43\% to 29\% in the realtime setting. Precision also falls drastically from 83\% to just 6.2\%---however, prior work has well documented the noisiness of realtime moderation decisions from Reddit communities. For example, Chandrasekharan et~al. found that for Crossmod, a realtime moderation system they designed, Reddit moderators would have moderated \textit{95\% of unmoderated comments} Crossmod flagged had the tool existed~\cite{chandrasekharan2019crossmod}, emphasizing that precision is a poor metric for evaluating realtime moderation tools on Reddit. Our results echo this idea, and highlight the need to evaluate systems \textit{in-context} (i.e., in collaboration with moderators) to understand the practical impact such systems might offer.

\subsection{Case Study: \texttt{r/worldnews}}
\begin{table}[t]
    \centering
    \small
    \begin{subtable}[b]{0.44\columnwidth}
        \begin{tabularx}{\columnwidth}{Xr}
            \toprule
            Reason     &       \%  \\
            \midrule
            Model Correct   &   48\% \\
            Mistaken Attack &   17\% \\
            Generalization &   14\% \\
            Hate Speech &   8\% \\
            Sarcasm &   8\% \\
            \bottomrule
        \end{tabularx}
        \subcaption{False pos.}
    \end{subtable}
    \begin{subtable}[b]{0.54\columnwidth}
        \begin{tabularx}{\columnwidth}{Xr}
            \toprule
            Reason     &       \% \\
            \midrule
            Missing Attack &   49\% \\
            Model Correct   &   16\% \\
            Sarcasm &  21\% \\
            Missing 3rd Person Atk  &   8\% \\
            Low Effort  &   4\% \\
            \bottomrule
        \end{tabularx}
        \subcaption{False neg.}
    \end{subtable}
    \caption{\textbf{Explaining errors in moderating \texttt{r/worldnews}}---%
        We show the most frequent explanations for errors generated by GPT-3.5 when moderating \texttt{r/worldnews}. We \textit{agree} with model decisions for 48\% of false positives and 16\% of false negatives. In general, models often make mistakes due to missing context.
    }
    \label{table:worldnews_explanations}
\end{table}

We complement our quantitative analyses with a qualitative analysis on errant decisions made by GPT-3.5 on the subreddit \texttt{r/worldnews}, which is a subreddit dedicated to major news around the world. We chose this particular subreddit as it is popular (33M subscribers), it well-balances false positives (33\%) and false negatives (67\%) to analyze, and tends to more general discussion versus niche subcommunity interests. \texttt{r/worldnews} has only 3~rules that apply to comments, two of which are related to personal attacks and hate speech, and one which is related to meme-content. Two independent raters manually investigated the 244~comments that produced either a false negative or false positive, iteratively developed a codebook for explaining each error, and coded each comment. Inter-rater agreement was high (Cohen's Kappa 0.89), and disagreements were manually resolved. Due to noisiness in subreddit moderation, we first determined whether Table~\ref{table:worldnews_explanations} shows a distribution of common explanations for false positives and false negatives produced by the model. We discuss each in detail below:

\paragraph{False positives} We observe that for 48\% of false positives, we \textit{agree} with the model decision over the human rater. For example, the comment ``Lol you're a fucking idiot'' was not flagged by human moderators, but clearly violates the second rule of \texttt{r/worldnews} and the Reddiquette. The existence of discrepancies like this is encouraging as it suggests LLMs can potentially be used as a double-check aid for human moderators who may inadvertently make mistakes. The next top reasons are when GPT-3.5 mistakenly identifies a comment as an attack (17\%), as a generalization (14\%), or as hate speech (8\%).
For example, the comment ``Are you aware that there are other businessmen in the world besides your Jew Boogeyman?'' was flagged as violative of hate speech rules, despite the fact that it is attempting to rebuff the original commenter and defend Jewish people from unwanted stereotypes. Given that hate speech and hateful language is part of the Reddiquette, errors produced by LLMs like these will have a wide impact on all communities wishing to curb and moderate hate speech.

\paragraph{False negatives} 49\% of false negatives were due to GPT-3.5 missing a personal or identity attack. It is not immediately clear why the model misses these: for example, the comment ``Not sure what that has to do with free speech you moron. Are you equating murder to speech?'' is not flagged, despite an obvious personal attack. In 8\% of other cases, the attack was indirect, for example: one comment begins with ``She'll never do time, don't be a retard...'' which, due to a mix of subjects in the sentence, may be hard to immediately discern. We agree with the model decision in 16\% of cases (i.e., it was unclear why a comment was moderated by human raters). In general, we observe that most errors (false positive or false negative) are due to some kind of \textit{missing context}: this could come from the conversation itself, from the post, or from some shared knowledge about the world. Such errors suggest that even with LLMs as an aid, human-review is still critical to address algorithmic mistakes.



\subsection{Does Conversational Context Reduce Errors?}
Given that a significant fraction of errors were made due to missing context, we next investigate how embedding conversational context into the prompt itself affects model performance. We began by collecting a corpus of 8859~errors made by each subreddit-specific LLM during our realtime experiment and, if the comment was in reply to another comment, collected the parent\_comment in the comment thread. This resulted in 4131 $(parent, child)$ comment pairs. We then prompted the LLM with a slightly modified prompt to account for the comment tree (included in the Appendix) and evaluated each comment pair. We found that adding the parent comment corrects 35\% of errors, thereby significantly improving the performance of each subreddit-specific LLM. Embedding context has a much higher impact on correcting false positives (40\% were corrected) than on correcting false negatives (6\% were corrected). Our results suggest that LLMs can be steered with appropriately provided conversation context for online moderation tasks, a task which has proven challenging for other moderative tasks with older families of models~\cite{pavlopoulos2020toxicity}.

\subsection{Case Study: Comparison to Gemini Pro}
\begin{table}
    \centering
    \small
    \begin{tabularx}{\columnwidth}{Xrrr}
    \toprule
    Subreddit   &   Metrics &   GPT-3.5 & Gemini Pro \\
    \midrule
    \multirow{3}{*}{r/movies}   &  Accuracy & 0.81 & 0.81 \\
                                &  Precision & 0.95 & 0.92 \\
                                &  Recall & 0.68 & 0.7 \\
                                &  F1 & 0.79 & 0.8 \\
    \midrule
    \multirow{3}{*}{r/IAmA}     & Accuracy & 0.64 & 0.62 \\
                                & Precision & 0.84 & 0.78 \\
                                & Recall & 0.37 & 0.38 \\
                                & F1 & 0.52 & 0.51 \\
    \midrule
    \multirow{3}{*}{r/askscience}   &   Accuracy & 0.38 & 0.4 \\
                                    &   Precision & 0.87 & 0.95 \\
                                    &   Recall & 0.11 & 0.11 \\
                                    &   F1 & 0.2 & 0.19 \\

    \bottomrule    
    \end{tabularx}
    \caption{\textbf{Rule-based moderation with GPT-3.5 and Gemini Pro}---%
        GPT-3.5 and Gemini Pro have near identical performance across the subreddits we tested, suggesting strong parity between the models in rule-based moderation tasks.
    }
    \label{table:reddit_gemini_comparison}
\end{table}

Finally, we compare our results from GPT-3.5 to another commodity, state-of-the-art LLM, Google's Gemini Pro. Despite limited details about model size, independent research has put Gemini Pro approximately on par with GPT-3.5 on common language benchmarks~\cite{akter2023depth}. As such, we view this case-study as an apt comparison to evaluate if two LLMs with completely different training data, hyperparameters, and system design decisions behave similarly in content moderation settings. To evaluate Gemini Pro, we evaluate the same three subreddits we examined when studying prompt variability with an identical prompt. Indeed, we observe that GPT-3.5 and Gemini Pro achieve near identical performance in all metrics, with GPT-3.5 slightly favoring precision where Gemini Pro slightly favors recall (Table~\ref{table:reddit_gemini_comparison}). These results are encouraging, as they show that our results generalize to other LLMs. Furthermore, community moderators wishing to deploy an LLM need not rely on a single model to aid in their moderation tasks.
\section{Toxicity Detection}
\label{sec:toxicity}

\begin{table*}
    \centering
    \begin{tabularx}{0.8\linewidth}{crrrrr}
    \toprule
    Model   &  Threshold   & Accuracy & Precision    & Recall   & F1  \\
    \midrule
    Perspective API \texttt{TOXICITY} & 0.5 & 0.66 & 0.68 & 0.61 & 0.64 \\
    Perspective API \texttt{TOXICITY} & 0.7 & 0.59 & 0.76 & 0.27 & 0.4 \\
    Perspective API \texttt{TOXICITY} & 0.9 & 0.52 & 0.92 & 0.05 & 0.1 \\
    Perspective API \texttt{SEVERE\_TOXICITY} & 0.5 & 0.51 & \textbf{0.96} & 0.02 & 0.05 \\
    OpenAI GPT-3 & -- & 0.71 & 0.71 & 0.73 & 0.72 \\
    OpenAI GPT-3.5 & 7/10 & \textbf{0.73} & 0.7 & 0.81 & \textbf{0.75} \\
    OpenAI GPT-4 & 5/10 & \textbf{0.73} & 0.71 & 0.78 & \textbf{0.75} \\
    Google Gemini Pro   & 8/10    &   0.7 &   0.64    &   \textbf{0.86}    &   0.74 \\
    \bottomrule
    \end{tabularx}
    \caption{\textbf{Toxicity detection aggregate performance}---%
    We show aggregate performance from the Perspective API's \texttt{TOXICITY} and \texttt{SEVERE\_TOXICITY} models at varying thresholds (0.5, 0,7, and 0.9) and the LLMs we evaluate. Across almost every metric, the LLMs we tested outperformed the Perspective API at toxicity detection. Surprisingly, we saw only marginal improvements at this task when comparing smaller LLMs (GPT-3) to large ones (GPT-4, Gemini Pro), suggesting a potential limit to LLM performance at toxicity detection.
    }
    \label{table:aggregate_performance}
\end{table*}

In this section, we investigate how well several commodity LLMs perform at toxicity detection when compared to existing baselines and identify opportunities to improve LLM performance for toxicity detection.


\subsection{Methods}
\subsubsection{Dataset}
We leverage a dataset of toxicity annotations curated by Kumar et~al.~\shortcite{kumar2021designing} that consists of 107K~comments sourced from three social platforms: Reddit, Twitter, and 4chan. Each comment was labeled for toxicity by five~raters on a 5-point Likert scale from ``not at all toxic'' to ``very toxic.'' We collapse this score into a binary outcome based on the median toxicity rating. If the median rating was equal to or above ``moderately toxic'' we denote the comment as toxic, and if the rating was below this threshold we denote the comment as non-toxic. Due to cost limitations, we do not evaluate each LLM on the entire dataset. Instead, we sample the aggregate dataset to build a balanced dataset of toxicity annotations that contains 5000~examples of toxic content and 5000~examples of nontoxic content.


\subsubsection{Prompt design}
We first prompt each LLM with the definition of toxic content used by Google Jigsaw, which is ``a rude, disrespectful, or unreasonable comment that is likely to make someone leave a discussion'' ~\cite{perspectiveapi}. We then ask the model to rate the toxicity of unseen comments on a scale from 1--10, as well as to provide an explanation for the rating in natural language. For ease of downstream analysis, we prompted the LLM to always return results as a JSON object with consistent indexing. The full prompt text is: 

\begin{small}
\begin{verbatim}
Consider the following comment: 
<example_comment>

Return a JSON object with a two fields, 
"rating," that is a score from 1 to 10 
depending on how toxic the comment is 
based on our definition, and "explanation" 
which contains a reason for the rating.    
\end{verbatim}
\end{small}

\subsubsection{Evaluation strategy}
We evaluate four state-of-the-art LLMs for this task---OpenAI's GPT3, OpenAI's GPT3.5 (i.e., ChatGPT), OpenAI's GPT4, and Google's Gemini Pro. As a case study, we also compare these results against Meta's open source LLM, Llama 2. We chose these as they represent recent, publicly available LLM offerings on the market and covers a range of model sizes and training strategies. When applicable by each API, we set the temperature to~0 to ensure determinism across repeated queries. To compare across a SOTA commodity baseline, we also label each comment using the Google Jigsaw Perspective API~\cite{perspectiveapi} using their \texttt{TOXICITY} and \texttt{SEVERE\_TOXICITY} classifiers at varying thresholds.

\subsection{Aggregate Performance}
The LLMs we study outperform the Perspective API at toxic content detection across almost all metrics, chiefly in accuracy and in balancing the precision-recall tradeoff (Table~\ref{table:aggregate_performance}). OpenAI's GPT models achieve an accuracy of 0.71--0.73, compared to a maximum accuracy of 0.66 from the Perspective API's TOXICITY classifier at a threshold of 0.5. Similarly, Google's Gemini Pro achieves an accuracy of 0.7 and achieves the greatest recall across all models we evaluate (0.86). The LLMs we evaluate well-balance precision and recall, achieving F1 scores of 0.72--0.75, compared to a maximum F1 of 0.64 from Perspective. Prior work noted that Perspective favors precision over recall~\cite{kumar2021designing}; we observe a similar trend here: the (\texttt{SEVERE\_TOXICITY}) model achieves near perfect precision (0.96) at high thresholds, but with a significant impact to recall (0.02). Interestingly, there is only marginal improvement on this particular toxicity detection task between GPT-3, GPT-3.5, GPT-4, and Gemini Pro, despite a significant increase in parameter size between the models. Although GPT-3.5 and GPT-4 return the highest accuracy and F1 scores (0.75), GPT-3's is comparable, achieving an F1 of 0.72. The lack of significant improvement with increased model size suggests a potential plateau at toxic content detection for larger models.


\subsection{Prompt Sensitivity}
\begin{table}
    \centering
    \small
    \begin{tabularx}{\columnwidth}{Xrrr}
        \toprule
        Metric      &   Base     &   Chain-of-Thought     &   No Definition \\
        \midrule
        Accuracy    &    \textbf{0.73}   & \textbf{0.73}          &   0.54 \\
        Precision   &    0.69   & 0.69          &   \textbf{0.85} \\
        Recall      &    \textbf{0.84}   & 0.81          &   0.13  \\
        \midrule
        F1          &    \textbf{0.75}   & 0.74          &   0.22 \\
        \bottomrule
    \end{tabularx}
    \caption{\textbf{GPT-3.5 Prompt Sensitivity with Toxic Content Detection}---%
        Toxicity detection is moderately sensitive to prompts: while chain-of-thought prompting
        produced nearly identical results to our base prompt, prompting with no definition saw a significant decrease
        in performance, highlighting the importance of adding proper definitions when prompting LLMs for classification.
    }
    \label{table:prompt_variance_toxicity}
\end{table}
Similarly to our community-based moderation evaluation, we examine how two other prompting strategies to see if our results are robust to prompt inputs: chain-of-thought prompting and prompting without a definition of toxicity. We evaluate each strategy on a subsample of our golden dataset consisting of 1000~comments (500~toxic and 500~nontoxic). Our prompting experiments are consistent across each LLM we evaluated; Table~\ref{table:prompt_variance_toxicity} shows the results for GPT-3.5 as an exemplar. We do not observe any differences between our ``base'' prompt and a prompt which elicits chain-of-thought reasoning. There is, however, a significant performance decrease when the prompt does not contain a definition of toxicity. Interestingly, the no-definition prompt produces \textit{fewer} false positives (resulting in a higher precision of 0.85 compared to 0.69), however, recall is dramatically reduced (from 0.84 to 0.12), highlighting the importance of providing sufficient world-context in order to use LLMs for any classification task.

\subsection{Thresholding LLMs for Toxicity Detection}
For many toxic content detection tasks, it is often useful to be able to ``threshold'' a toxic content classifier towards higher recall or higher precision depending on the use-case. For example, prior research has favored a high precision signal when investigating toxic behaviors online~\cite{kumar2023understanding} and used a high threshold for detection, whereas other research favored F1 for a balance between precision and recall~\cite{saveski2021structure} and used a lower threshold for detection. Although LLMs are not explicitly designed as classifiers, we can simulate a classification-like thresholding by selecting a threshold from the score produced by the LLM (1--10).

\begin{figure}[ht]
    \centering
    \includegraphics[width=\columnwidth]{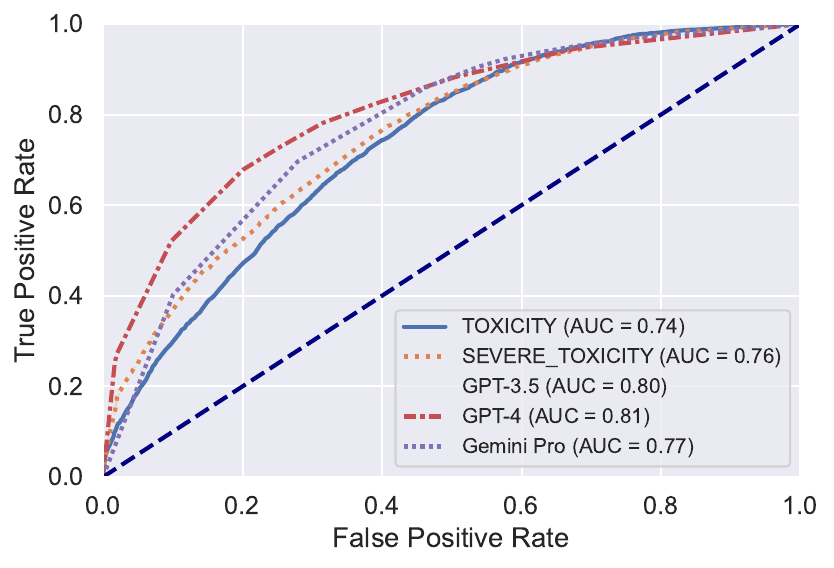}
    \caption{\textbf{Toxicity detection ROC curves}---%
            We show the ROC curves for all toxicity detection systems we evaluate. LLMs from OpenAI and Google outperform the Perspective API in AUC, achieving an AUC of 0.77--0.81 compared to a maximum AUC of 0.76 from the Perspective API.
    }
    \label{fig:roc_auc}
\end{figure}

We observe that varying the thresholds for LLMs \textit{does} produce an internally consistent precision/recall tradeoff for GPT-3.5, GPT-4, and Gemini Pro. Figure~\ref{fig:roc_auc} shows the resultant receiver operating characteristic curve for each model we evaluate. The area under the ROC curve (ROC\_AUC) for GPT-3.5, GPT-4, and Gemini Pro \textit{surpasses} the AUC of the Perspective API models (0.77--0.81 compared to 0.76), demonstrating that the LLMs are better able to balance precision and recall better than Perspective models can. As such, prompting LLMs to provide a discrete rating for toxicity can be another useful lever to content moderators looking to curb toxic content.


\begin{table*}
    \centering
    \small
    \begin{tabularx}{0.67\linewidth}{X|rrr|rrr|rrr}
        \toprule
        Threshold   &   \multicolumn{3}{c|}{GPT-3.5} &   \multicolumn{3}{c|}{GPT-4}  &   \multicolumn{3}{c}{Gemini Pro} \\
        \midrule
                    &   Prec.    &   Recall  &   F1  &   Prec    &   Recall  &   F1 &   Prec    &   Recall  &   F1 \\
        \midrule
        2 & 0.53 & \textbf{0.98} & 0.69 & 0.58 & \textbf{0.95} & 0.72 & 0.55 & \textbf{0.97} & 0.7 \\
        3 & 0.61 & 0.92 & 0.73 & 0.64 & 0.88 & 0.74 & 0.57 & 0.96 & 0.72 \\
        4 & 0.67 & 0.87 & \textbf{0.75} & 0.68 & 0.83 & 0.74 & 0.58 & 0.95 & 0.72 \\
        5 & 0.68 & 0.84 & \textbf{0.75} & 0.71 & 0.78 & \textbf{0.75} & 0.59 & 0.94 & 0.73 \\
        6 & 0.7 & 0.81 & \textbf{0.75} & 0.74 & 0.73 & 0.74 & 0.61 & 0.92 & 0.73 \\
        7 & 0.74 & 0.74 & 0.74 & 0.77 & 0.68 & 0.72 & 0.62 & 0.91 & \textbf{0.74} \\
        8 & 0.8 & 0.58 & 0.67 & 0.84 & 0.52 & 0.64 & 0.64 & 0.86 & \textbf{0.74} \\
        9 & 0.88 & 0.34 & 0.49 & 0.93 & 0.27 & 0.42 & 0.71 & 0.7 & 0.7 \\
        10 & \textbf{0.88} & 0.34 & 0.49 & \textbf{0.94} & 0.25 & 0.39 & \textbf{0.8} & 0.4 & 0.53 \\
        \bottomrule
    \end{tabularx}
    \caption{\textbf{Thresholding LLMs for toxicity detection}---%
        GPT-3.5, GPT-4, and Gemini Pro all produce an internally consistent precision/recall tradeoff when performing toxicity detection, highlighting how such models can serve similar functions to toxicity detection classifiers.
    }
    \label{table:threshold_analysis}
\end{table*}

Indeed, varying classification thresholds does meaningfully impact precision, recall, and F1 performance for the LLMs we tested (Table~\ref{table:threshold_analysis}). At the highest thresholds, GPT-4 is able to achieve comparable precision (0.94) to the Perspective API (0.96), but without as a significant hit to recall (0.25 vs. 0.02). Gemini Pro and GPT-3.5 are not able to achieve similarly high precision as GPT-4 or Perspective's \texttt{SEVERE\_TOXICITY} classifier (0.8--0.88), however, both well-balance recall with precision (F1 of 0.49--0.53, t=10), highlighting that these LLMs can be equally effective at detecting toxic content in settings where balancing precision and recall is critical.

\subsection{Do LLM Toxicity Ratings Align with Human Raters?}
Although many use-cases of toxicity classifiers in content moderation settings are binary, the \textit{severity} of the comment also plays a role in downstream moderation action. Recent research has demonstrated that LLMs are comparable to crowdsourced human-raters at many tasks~\cite{tornberg2023chatgpt,gilardi2023chatgpt}, however, this has yet to be tested in a toxicity detection setting. 

To measure how well LLM toxicity ratings align with human-raters, we compute a Pearson correlation between the rating provided by the LLM (from 1--10) and the median toxicity score across all five raters for each comment (from 1--5) for all comments labeled by GPT-3.5, GPT-4, and Gemini Pro. Intuitively, this gives a sense of how closely the LLM~produced score aligns with raters, and if higher scores correspond to higher levels of severity. We find there is a strong correlation (GPT-3.5, $r=0.60$, GPT-4, $r=0.61$) between GPT-3.5, GPT-4, and human raters, suggesting that such ratings can also provide a rough proxy for the severity of toxic comments for end-users. There is a weaker correlation between Gemini Pro and human raters ($r=0.5$). For downstream moderation tasks, nuanced notions of severity of content can serve as an additional signal to enable automated, \textit{multilayered} content moderation, wherein the severity of the content is taken into consideration when deciding the moderation action to take~\cite{multilayered}.

\begin{table}[t]
    \centering
    \small
    \begin{subtable}[b]{0.49\columnwidth}
        \begin{tabularx}{\columnwidth}{Xr}
            \toprule
            Tag     &       \% \\
            \midrule
            Language        &   34\% \\
            Model Correct   &   28\% \\
            Tone            &   13\% \\
            Sexual          &   10\% \\
            Stereotype      &   8\% \\
            \bottomrule
        \end{tabularx}
        \subcaption{False pos.}
    \end{subtable}
    \begin{subtable}[b]{0.49\columnwidth}
        \begin{tabularx}{\columnwidth}{Xr}
            \toprule
            Tag     &       \% \\
            \midrule
            Missing Attack      &   45\% \\
            Sarcasm / Humor     &   23\% \\
            Opinions        &   22\% \\
            Neutral         &   7\% \\
            \\
            \bottomrule
        \end{tabularx}
        \subcaption{False neg.}
    \end{subtable}
    \caption{\textbf{Explaining errors in toxicity detection}---%
        We show the most common explanations for errors generated by GPT-4. False positives are largely due to misidentifying targets of harassment and offensive language, whereas false negatives are caused by a failure to identify implicitly hateful comments and an overreliance on opinions.  
    }
    \label{table:tox_explanations}
\end{table}

\subsection{Where Do LLMs Make Mistakes?}
To qualitatively investigate errors that exist when using LLMs for toxic content detection, we randomly sample 200~comment decision \emph{mistakes} from GPT-4: 100~of which are false positives (e.g., erroneously flagging a comment as toxic) and 100~of which are false negatives (e.g., failing to catch a toxic comment.) Two independent raters manually rated each comment, developed a codebook for each error, and coded each comment. Inter-rater agreement was high (Cohen's Kappa 0.85) and disagreements were manually resolved. Table~\ref{table:tox_explanations} shows the distribution of explanations provided for false positives and false negatives in our sample. We describe each class of errors below:

\paragraph{False positives} 
The primary reason LLMs erroneously flag benign content is due to triggering on poor language (e.g., profanity, slurs) and stereotypes (34\%), even in instances of neutral or positive connotation. For example, the comment ``Yo n***a wears jean shorts'' was not found to be toxic by human raters, but the model flagged this comment for ``using derogatory racial slurs.'' Prior work has highlighted how toxicity models can inherit unintended social biases~\cite{dixon2018measuring,halevy2021mitigating}---such considerations should also apply when leveraging LLMs for content moderation focused tasks. In many cases (47\%), GPT-4 erroneously identifies subjects of a comment as targets of the harassment: for example, a comment that discussed the coaching of a sports team (``Not the players bro.. Buh [sic] foolish coach'') was flagged as toxic due to the \textit{coach} finding the comment hurtful, despite the fact that the coach of the team was not a part of the online conversation. Similarly to our results with rule-based moderation, we find we agree with the model's decision in 28\% of cases.

\paragraph{False negatives}
The top reason that LLMs miss toxic content flagged by human raters is that the comment contains no explicit or direct threat or insult to another individual (27\%). For example, the comment: ``There is hunger and illiteracy in the land. Only in Nigeria.'' was not flagged by the model, due to there being no personal attacks. However, the comment still generalizes about the country of Nigeria, and human raters found this comment to be toxic. The model was reticent to flag what it deems to be sarcasm or humor as toxic (23\%), in contrast to human raters who often deemed such messages as toxic. Similarly, the model is also reticent to flag what it deems to be opinions as toxic (22\%), especially when the opinion is couched in a question. For example, the comment ``They offered the cheapest goods for a long time in the US. Why do you think that's not beneficial?'' was not flagged as toxic, despite prior work identifying that rhetorical tone can lead to increased toxicity and conversational collapse~\cite{zhang2018conversations}.

\subsection{Case Study: Comparison to Llama 2}
Finally, as a case study, we compare our results to Llama 2: an open-source foundation model created and distributed by Meta~\cite{touvron2023llama}. Llama 2 is an interesting comparison point as the family of models can be deployed on commodity GPUs, thus greatly reducing the cost of using an industry-grade LLM in practice. For our experiments, we focus on the 13B~parameter Llama 2 model, as it balances ease-of-deployment with reasonable model size. We ran every comment in our golden dataset through the model, prompting it identically to how we prompted other LLMs. Unfortunately, we observe that the model is difficult to steer properly. Frequent hallucinations greatly reduce capacity for consistent output, with 56\% of responses containing no valid JSON result. Even when the model does produce what appears to be JSON output, it is often improperly encoded or escaped (12\% of cases). Only 32\% of responses contained valid, parseable data; the model made predictions on 1936 (51\%) nontoxic comments and 1831 (49\%) toxic responses.

For valid responses, Llama 2 achieves decent performance: it reaches a maximum F1~score of 0.68, with an accuracy of 63\% and a precision of 0.59. These results are significantly worse than other LLMs we evaluated, albeit with the tradeoff that the model can run locally. Even at the highest threshold values, Llama 2 only achieves a maximum precision of 0.79, which is comparable to average performance from larger LLMs. While the promise of deploying an open source, on-device LLM is high, our results suggest that at present, Llama 2 cannot adequately tackle toxic content detection for content moderators. 

\section{Discussion}
Our work studied how well LLMs perform on two content-moderation related tasks: rule-based community moderation and toxic content detection. Our results provide insight in where existing LLMs excel and informs what gaps need to be addressed for such tools to have widespread impact in content moderation contexts. We now discuss our results, outline future directions for research in LLM-powered content moderation, and highlight limitations of our approach.

\paragraph{Increasing transparency in LLM decisions}
LLM performance can rapidly change over time, as we observed in our rule-based moderation experiments. The root cause of why model changes impact moderation decisions are unclear, but our results align with prior work in studying longitudinal LLM behavior inconsistencies~\cite{chen2023chatgpt}. Content moderators seeking to incorporate LLMs must exercise caution when deploying such tools in practice to maintain the reliability and stability of their moderation decisions. One area of future work is thus to identify tasks that change drastically when models change, to develop better benchmarks for continuous monitoring of LLMs.

\paragraph{Improving rule-based moderation with behavior and context}
Rule-based LLM performance falls below existing AI-assisted content moderation tools. For example, Chandrasekharan et~al.'s CrossMod tool~\shortcite{chandrasekharan2019crossmod}, which is developed using the same dataset as the one we leveraged in this paper, is able to achieve an accuracy of 86\%, compared to GPT-3.5's 64\%. One potential reason for this is that while LLMs are able to reason in a ``forward direction,'' by interpreting rules and examining if content falls within those rules, CrossMod is a ``reverse direction'' tool which starts from actual content removals and learns patterns that can be applied to other communities. An area of future work that might be promising is to see if combining rule-based reasoning with few-shot prompting behavioral examples might improve overall LLM performance without requiring any additional training or fine-tuning of the underlying models.

Another promising direction to improve rule-based moderation is in incorporating conversational or world context into decision making. We observed how incorporating context can correct 35\% of errors with little addition to the prompt, and future work can investigate incorporating other types of context (e.g., post details or images in a multimodal context) to aid in improving moderation outcomes.

\paragraph{Performance saturation in toxicity detection}
Although LLMs outperform existing commercial toxicity detection tools, we observed that LLM performance only marginally improves at toxicity detection despite the fact that newer model sizes are increasing by orders of magnitude~\cite{gpt4-model-size}. As such, our results paint caution to the notion that increasing model sizes will necessarily lead to downstream improvements in content moderation tasks. One potential area of future work can be in identifying what aspects of toxicity detection remain challenging \textit{in spite of} using an LLM (e.g., implicit hate or identity language) and potentially offload gray area comments to either more specialized automated pipelines or human review. 

\paragraph{Future experiments in content moderation}
Given that our LLMs performed better in a balanced context than in a real-world context, a promising area of future work would be to work collaboratively with moderators to see if LLMs might improve their existing moderation workflows. LLMs might help to steer moderators to potentially violative content, especially given our observation that ~48\% of false positives seem to violate community rules. Furthermore, moderators might have sufficient world-context to help steer LLMs through techniques like few-shot prompting~\cite{brown2020language} or through structure conversation~\cite{argyle2023leveraging}.

\paragraph{Limitations of our work}
This work primarily focuses on text-based content moderation tasks, and specifically on comment moderation on discussion oriented platforms. As such, our results may not extend to other types of moderation, for example, those involving mixed modalities (i.e., images, videos, etc.) or those on social platforms with other affordances (i.e., real-time communication). Furthermore, our evaluation here is primarily done on \textit{balanced datasets}---our results with examining real-world distributions suggest that significantly more work needs to be done to evaluate such tools in practice. We also note that while are our results are promising, the LLMs we tested are \textit{too expensive} to immediately impact most content moderation contexts, and future research is needed to build models that can balance performance with cost.

\section{Conclusion} 
Overall, our results provide a tempered but optimistic view of introducing LLMs into content-moderation contexts. We hope that our work provides the groundwork for researchers and practitioners and charts a pathway forward for more work in studying how LLMs can shape and impact the future of content moderation. We plan to release our code, prompts, and data at publication time to enable further research on automated content moderation with LLMs.


\bibliography{main}

\begin{thebibliography}{43}
\providecommand{\natexlab}[1]{#1}

\bibitem[{Akter et~al.(2023)Akter, Yu, Muhamed, Ou, B{\"a}uerle, Cabrera, Dholakia, Xiong, and Neubig}]{akter2023depth}
Akter, S.~N.; Yu, Z.; Muhamed, A.; Ou, T.; B{\"a}uerle, A.; Cabrera, {\'A}.~A.; Dholakia, K.; Xiong, C.; and Neubig, G. 2023.
\newblock An In-depth Look at Gemini's Language Abilities.
\newblock \emph{arXiv preprint arXiv:2312.11444}.

\bibitem[{Argyle et~al.(2023)Argyle, Bail, Busby, Gubler, Howe, Rytting, Sorensen, and Wingate}]{argyle2023leveraging}
Argyle, L.~P.; Bail, C.~A.; Busby, E.~C.; Gubler, J.~R.; Howe, T.; Rytting, C.; Sorensen, T.; and Wingate, D. 2023.
\newblock Leveraging AI for democratic discourse: Chat interventions can improve online political conversations at scale.
\newblock \emph{Proceedings of the National Academy of Sciences}.

\bibitem[{Atreja et~al.(2023)Atreja, Im, Resnick, and Hemphill}]{atreja2023appealmod}
Atreja, S.; Im, J.; Resnick, P.; and Hemphill, L. 2023.
\newblock AppealMod: Shifting Effort from Moderators to Users Making Appeals.
\newblock \emph{arXiv preprint arXiv:2301.07163}.

\bibitem[{Bastian(2023)}]{gpt4-model-size}
Bastian, M. 2023.
\newblock \url{https://the-decoder.com/gpt-4-has-a-trillion-parameters}.

\bibitem[{Brown et~al.(2020)Brown, Mann, Ryder, Subbiah, Kaplan, Dhariwal, Neelakantan, Shyam, Sastry, Askell et~al.}]{brown2020language}
Brown, T.; Mann, B.; Ryder, N.; Subbiah, M.; Kaplan, J.~D.; Dhariwal, P.; Neelakantan, A.; Shyam, P.; Sastry, G.; Askell, A.; et~al. 2020.
\newblock Language models are few-shot learners.
\newblock \emph{Advances in neural information processing systems}.

\bibitem[{Chandrasekharan et~al.(2019)Chandrasekharan, Gandhi, Mustelier, and Gilbert}]{chandrasekharan2019crossmod}
Chandrasekharan, E.; Gandhi, C.; Mustelier, M.~W.; and Gilbert, E. 2019.
\newblock Crossmod: A cross-community learning-based system to assist reddit moderators.
\newblock In \emph{ACM CSCW}.

\bibitem[{Chandrasekharan et~al.(2018)Chandrasekharan, Samory, Jhaver, Charvat, Bruckman, Lampe, Eisenstein, and Gilbert}]{chandrasekharan2018internet}
Chandrasekharan, E.; Samory, M.; Jhaver, S.; Charvat, H.; Bruckman, A.; Lampe, C.; Eisenstein, J.; and Gilbert, E. 2018.
\newblock The Internet's hidden rules: An empirical study of Reddit norm violations at micro, meso, and macro scales.
\newblock In \emph{ACM CSCW}.

\bibitem[{Chen, Zaharia, and Zou(2023)}]{chen2023chatgpt}
Chen, L.; Zaharia, M.; and Zou, J. 2023.
\newblock How is ChatGPT's behavior changing over time?
\newblock \emph{arXiv preprint arXiv:2307.09009}.

\bibitem[{Chiu, Collins, and Alexander(2021)}]{chiu2021detecting}
Chiu, K.-L.; Collins, A.; and Alexander, R. 2021.
\newblock Detecting hate speech with gpt-3.
\newblock \emph{arXiv preprint arXiv:2103.12407}.

\bibitem[{Choi and Lee(2023)}]{choi2023creator}
Choi, Y.; and Lee, M.~K. 2023.
\newblock Creator-friendly Algorithms: Behaviors, Challenges, and Design Opportunities in Algorithmic Platforms.
\newblock In \emph{ACM Conference on Human Computer Interaction}.

\bibitem[{Dixon et~al.(2018)Dixon, Li, Sorensen, Thain, and Vasserman}]{dixon2018measuring}
Dixon, L.; Li, J.; Sorensen, J.; Thain, N.; and Vasserman, L. 2018.
\newblock Measuring and mitigating unintended bias in text classification.
\newblock In \emph{AAAI Conf.\ on AI, Ethics, and Society}.

\bibitem[{ElSherief et~al.(2021)ElSherief, Ziems, Muchlinski, Anupindi, Seybolt, De~Choudhury, and Yang}]{elsherief2021latent}
ElSherief, M.; Ziems, C.; Muchlinski, D.; Anupindi, V.; Seybolt, J.; De~Choudhury, M.; and Yang, D. 2021.
\newblock Latent hatred: A benchmark for understanding implicit hate speech.
\newblock \emph{arXiv preprint arXiv:2109.05322}.

\bibitem[{Fiesler et~al.(2018)Fiesler, Jiang, McCann, Frye, and Brubaker}]{fiesler2018reddit}
Fiesler, C.; Jiang, J.; McCann, J.; Frye, K.; and Brubaker, J. 2018.
\newblock Reddit rules! characterizing an ecosystem of governance.
\newblock In \emph{Int.\ AAAI Conference on Web and Social Media}.

\bibitem[{Franco, Gaggi, and Palazzi(2023)}]{franco2023analyzing}
Franco, M.; Gaggi, O.; and Palazzi, C.~E. 2023.
\newblock Analyzing the Use of Large Language Models for Content Moderation with ChatGPT Examples.
\newblock In \emph{International Workshop on Open Challenges in Online Social Networks}.

\bibitem[{Gilardi, Alizadeh, and Kubli(2023)}]{gilardi2023chatgpt}
Gilardi, F.; Alizadeh, M.; and Kubli, M. 2023.
\newblock Chatgpt outperforms crowd-workers for text-annotation tasks.
\newblock \emph{arXiv preprint arXiv:2303.15056}.

\bibitem[{Gilbert(2020)}]{gilbert2020run}
Gilbert, S.~A. 2020.
\newblock "I run the world's largest historical outreach project and it's on a cesspool of a website." Moderating a Public Scholarship Site on Reddit: A Case Study of r/AskHistorians.
\newblock \emph{ACM CSCW}.

\bibitem[{Gillespie(2020)}]{gillespie2020content}
Gillespie, T. 2020.
\newblock Content moderation, AI, and the question of scale.
\newblock \emph{Big Data \& Society}.

\bibitem[{{Google Jigsaw}(2018)}]{perspectiveapi}
{Google Jigsaw}. 2018.
\newblock Perspective API.
\newblock \url{https://www.perspectiveapi.com/}.

\bibitem[{Halevy et~al.(2021)Halevy, Harris, Bruckman, Yang, and Howard}]{halevy2021mitigating}
Halevy, M.; Harris, C.; Bruckman, A.; Yang, D.; and Howard, A. 2021.
\newblock Mitigating racial biases in toxic language detection with an equity-based ensemble framework.
\newblock In \emph{Equity and Access in Algorithms, Mechanisms, and Optimization}.

\bibitem[{Han et~al.(2023)Han, Seering, Kumar, Hancock, and Durumeric}]{han2023hate}
Han, C.; Seering, J.; Kumar, D.; Hancock, J.~T.; and Durumeric, Z. 2023.
\newblock Hate raids on Twitch: Echoes of the past, new modalities, and implications for platform governance.
\newblock In \emph{ACM CSCW}.

\bibitem[{Hanley and Durumeric(2023)}]{hanley2023twits}
Hanley, H.~W.; and Durumeric, Z. 2023.
\newblock Twits, Toxic Tweets, and Tribal Tendencies: Trends in Politically Polarized Posts on Twitter.
\newblock \emph{arXiv preprint arXiv:2307.10349}.

\bibitem[{He et~al.(2024)He, Zannettou, Shen, and Zhang}]{he2024prompt}
He, X.; Zannettou, S.; Shen, Y.; and Zhang, Y. 2024.
\newblock You Only Prompt Once: On the Capabilities of Prompt Learning on Large Language Models to Tackle Toxic Content.
\newblock In \emph{IEEE Symposium on Security and Privacy}.

\bibitem[{Jhaver, Bruckman, and Gilbert(2019)}]{jhaver2019does}
Jhaver, S.; Bruckman, A.; and Gilbert, E. 2019.
\newblock Does transparency in moderation really matter? User behavior after content removal explanations on reddit.
\newblock In \emph{ACM CSCW}.

\bibitem[{Jiang et~al.(2023)Jiang, Nie, Brubaker, and Fiesler}]{jiang2023trade}
Jiang, J.~A.; Nie, P.; Brubaker, J.~R.; and Fiesler, C. 2023.
\newblock A trade-off-centered framework of content moderation.
\newblock \emph{ACM Transactions on Computer-Human Interaction}.

\bibitem[{Kumar et~al.(2023)Kumar, Hancock, Thomas, and Durumeric}]{kumar2023understanding}
Kumar, D.; Hancock, J.; Thomas, K.; and Durumeric, Z. 2023.
\newblock Understanding the behaviors of toxic accounts on reddit.
\newblock In \emph{ACM Web Conference}.

\bibitem[{Kumar et~al.(2021)Kumar, Kelley, Consolvo, Mason, Bursztein, Durumeric, Thomas, and Bailey}]{kumar2021designing}
Kumar, D.; Kelley, P.~G.; Consolvo, S.; Mason, J.; Bursztein, E.; Durumeric, Z.; Thomas, K.; and Bailey, M. 2021.
\newblock Designing Toxic Content Classification for a Diversity of Perspectives.
\newblock In \emph{USENIX SOUPS}.

\bibitem[{Kuo, Hernani, and Grossklags(2023)}]{kuo2023unsung}
Kuo, T.; Hernani, A.; and Grossklags, J. 2023.
\newblock The Unsung Heroes of Facebook Groups Moderation: A Case Study of Moderation Practices and Tools.
\newblock In \emph{ACM CSCW}.

\bibitem[{Lai et~al.(2022)Lai, Carton, Bhatnagar, Liao, Zhang, and Tan}]{lai2022human}
Lai, V.; Carton, S.; Bhatnagar, R.; Liao, Q.~V.; Zhang, Y.; and Tan, C. 2022.
\newblock Human-ai collaboration via conditional delegation: A case study of content moderation.
\newblock In \emph{ACM Conference on Human Computer Interaction}.

\bibitem[{Lambert, Rajagopal, and Chandrasekharan(2022)}]{lambert2022conversational}
Lambert, C.; Rajagopal, A.; and Chandrasekharan, E. 2022.
\newblock Conversational Resilience: Quantifying and Predicting Conversational Outcomes Following Adverse Events.
\newblock In \emph{AAAI Intl.\ Conference on Web and Social Media}.

\bibitem[{Li et~al.(2023)Li, Fan, Atreja, and Hemphill}]{li2023hot}
Li, L.; Fan, L.; Atreja, S.; and Hemphill, L. 2023.
\newblock " HOT" ChatGPT: The promise of ChatGPT in detecting and discriminating hateful, offensive, and toxic comments on social media.
\newblock \emph{arXiv preprint arXiv:2304.10619}.

\bibitem[{Ma and Kou(2023)}]{ma2023defaulting}
Ma, R.; and Kou, Y. 2023.
\newblock Defaulting to boilerplate answers, they didn't engage in a genuine conversation: Dimensions of Transparency Design in Creator Moderation.
\newblock In \emph{ACM Conference on Computer-Supported Cooperative Work and Social Computing}.

\bibitem[{Oliveira et~al.(2023)Oliveira, Cecote, Silva, Gertrudes, Freitas, and Luz}]{oliveira2023good}
Oliveira, A.~S.; Cecote, T.~C.; Silva, P.~H.; Gertrudes, J.~C.; Freitas, V.~L.; and Luz, E.~J. 2023.
\newblock How Good Is ChatGPT For Detecting Hate Speech In Portuguese?
\newblock In \emph{Anais do XIV Simp{\'o}sio Brasileiro de Tecnologia da Informa{\c{c}}{\~a}o e da Linguagem Humana}. SBC.

\bibitem[{OpenAI(2023)}]{openai-prompt-guides}
OpenAI. 2023.
\newblock Prompt Engineering.
\newblock \url{https://platform.openai.com/docs/guides/prompt-engineering/prompt-engineering}.

\bibitem[{OpenWeb(2021)}]{multilayered}
OpenWeb. 2021.
\newblock How Multi-Layered Moderation Influences Positive Commenting.
\newblock \url{https://www.openweb.com/blog/how-multi-layered-moderation-influences-positive-commenting}.

\bibitem[{Pavlopoulos et~al.(2020)Pavlopoulos, Sorensen, Dixon, Thain, and Androutsopoulos}]{pavlopoulos2020toxicity}
Pavlopoulos, J.; Sorensen, J.; Dixon, L.; Thain, N.; and Androutsopoulos, I. 2020.
\newblock Toxicity Detection: Does Context Really Matter?
\newblock In \emph{Annual Meeting of the Association for Computational Linguistics}.

\bibitem[{Saveski, Roy, and Roy(2021)}]{saveski2021structure}
Saveski, M.; Roy, B.; and Roy, D. 2021.
\newblock The structure of toxic conversations on Twitter.
\newblock In \emph{ACM Web Conference}.

\bibitem[{Schick, Udupa, and Sch{\"u}tze(2021)}]{schick2021self}
Schick, T.; Udupa, S.; and Sch{\"u}tze, H. 2021.
\newblock Self-diagnosis and self-debiasing: A proposal for reducing corpus-based bias in nlp.
\newblock \emph{Transactions of the Association for Computational Linguistics}.

\bibitem[{T{\"o}rnberg(2023)}]{tornberg2023chatgpt}
T{\"o}rnberg, P. 2023.
\newblock Chatgpt-4 outperforms experts and crowd workers in annotating political twitter messages with zero-shot learning.
\newblock \emph{arXiv preprint arXiv:2304.06588}.

\bibitem[{Touvron et~al.(2023)Touvron, Martin, Stone, Albert, Almahairi, Babaei, Bashlykov, Batra, Bhargava, Bhosale et~al.}]{touvron2023llama}
Touvron, H.; Martin, L.; Stone, K.; Albert, P.; Almahairi, A.; Babaei, Y.; Bashlykov, N.; Batra, S.; Bhargava, P.; Bhosale, S.; et~al. 2023.
\newblock Llama 2: Open foundation and fine-tuned chat models.
\newblock \emph{arXiv preprint arXiv:2307.09288}.

\bibitem[{Wei et~al.(2022)Wei, Wang, Schuurmans, Bosma, Xia, Chi, Le, Zhou et~al.}]{wei2022chain}
Wei, J.; Wang, X.; Schuurmans, D.; Bosma, M.; Xia, F.; Chi, E.; Le, Q.~V.; Zhou, D.; et~al. 2022.
\newblock Chain-of-thought prompting elicits reasoning in large language models.
\newblock \emph{Advances in Neural Information Processing Systems}.

\bibitem[{White et~al.(2023)White, Fu, Hays, Sandborn, Olea, Gilbert, Elnashar, Spencer-Smith, and Schmidt}]{white2023prompt}
White, J.; Fu, Q.; Hays, S.; Sandborn, M.; Olea, C.; Gilbert, H.; Elnashar, A.; Spencer-Smith, J.; and Schmidt, D.~C. 2023.
\newblock A prompt pattern catalog to enhance prompt engineering with chatgpt.
\newblock \emph{arXiv preprint arXiv:2302.11382}.

\bibitem[{Xia et~al.(2020)Xia, Zhu, Lu, Zhang, and Gu}]{xia2020exploring}
Xia, Y.; Zhu, H.; Lu, T.; Zhang, P.; and Gu, N. 2020.
\newblock Exploring antecedents and consequences of toxicity in online discussions: A case study on reddit.
\newblock \emph{ACM Conference on Human Computer Interaction}.

\bibitem[{Zhang et~al.(2018)Zhang, Chang, Danescu-Niculescu-Mizil, Dixon, Hua, Thain, and Taraborelli}]{zhang2018conversations}
Zhang, J.; Chang, J.~P.; Danescu-Niculescu-Mizil, C.; Dixon, L.; Hua, Y.; Thain, N.; and Taraborelli, D. 2018.
\newblock Conversations gone awry: Detecting early signs of conversational failure.
\newblock \emph{arXiv preprint arXiv:1805.05345}.

\end{thebibliography}
\section{Appendix}
\subsection{Rule Type Classifier Details}
In order to add more descriptive labels to each rule in our dataset, we replicate the results from Fiesler~et al.~\shortcite{fiesler2018reddit}. We build three logistic regression classifiers for identifying restrictive rules, prescriptive rules, and format rules, as these types of rules were abundant in the training data. We use their training dataset and closely follow the model presented in their paper, which is a logistic regression classifier based on unigram, bigram, and trigram features extracted from each rule text. We held out 20\% of rules as a testing set with a random seed of~42. Ultimately, we were able to achieve comparable results to their paper: the restrictive model achieved an F1 of 0.93, the prescriptive model achieved an F1 of 0.89, and the format model achieved an F1 of 0.79.

\subsection{Conversational Context Prompt}

\begin{small}
\begin{verbatim}
Consider the following comment tree:
<comment_1> : "<example_comment>"
<comment_2> : "<example_comment_2>"

Return a JSON object with five fields, 
"would_moderate," that is 
either "yes" or "no" depending 
on if you would remove <comment_2>
from the subreddit, "rules" which are 
the text of the rules being violated, 
"rule_nums" which are a comma-separated
list of rules being violated, "rating"
which is a score from 1-5 on how 
violative the comment is, and "explanation"
which provides a reason for your decision.
\end{verbatim}
\end{small}

\subsection{Cost and Performance of LLMs for Moderation}

\begin{table}
    \centering
    \small
    \begin{tabularx}{\columnwidth}{Xrrr}
    \toprule
    Subreddit   &   Metrics &   Toxic Task & Rule Task \\
    \midrule
    \multirow{2}{*}{GPT-3}      &  Time taken & 3 days & -- \\
                                &  Cost & \$97 & -- \\
    \midrule
    \multirow{2}{*}{GPT-3.5}    &  Time taken & $<$1 day & 5 days \\
                                &  Cost & \$50 & \$175 \\
    \midrule
    \multirow{2}{*}{GPT-4}     & Time taken & 2 days & -- \\
                                & Cost & \$154 & -- \\
    \midrule
    \multirow{2}{*}{Gemini Pro}     &   Time taken & $<$1 day & 5 days* \\
                                    &   Cost & N/A & N/A \\

    \bottomrule    
    \end{tabularx}
    \caption{\textbf{Comparative cost and performance}---%
        We show each model's approximate time taken and the cost of each experiment. 
    }
    \label{table:cost_comparision}
\end{table}

Table~\ref{table:cost_comparision} shows a comparison between costs and overall ``performance'' (e.g., time taken to conduct the experiments) for each
model we evaluate. We stress that such estimates, especially time taken, are challenging to reproduce as they are impacted by factors often out of our control, such as model load.

\end{document}